\documentclass[letterpaper,twocolumn,prl,aps,showpacs,superscriptaddress,floatfix]{revtex4-1}
\usepackage[latin1]{inputenc}
\usepackage{bm}
\usepackage[usenames]{color}
\usepackage{multirow}
\usepackage{amssymb}
\usepackage{amsbsy}
\usepackage{amsmath}
\usepackage{stmaryrd}
\usepackage{graphicx}
\usepackage{epsfig}
\usepackage{placeins}
\usepackage{ulem}
\makeatletter

\begin{document}
\title{Heat Conductivity of the Heisenberg Spin-$1/2$ Ladder:\\ From Weak to Strong Breaking of Integrability}

\author{Robin Steinigeweg}
\email{rsteinig@uos.de}
\affiliation{Department of Physics, University of Osnabr\"uck, D-49069 Osnabr\"uck, Germany}
\affiliation{Institute for Theoretical Physics, Technical University Braunschweig, D-38106 Braunschweig, Germany}

\author{Jacek Herbrych}
\email{jacek@physics.uoc.gr}
\affiliation{Department of Physics, University of Crete, GR-71003 Heraklion, Greece}
\affiliation{Cretan Center for Quantum Complexity and Nanotechnology, University of Crete, GR-71003 Heraklion, Greece}

\author{Xenophon Zotos}
\email{zotos@physics.uoc.gr}
\affiliation{Department of Physics, University of Crete, GR-71003 Heraklion, Greece}
\affiliation{Cretan Center for Quantum Complexity and Nanotechnology, University of Crete, GR-71003 Heraklion, Greece}
\affiliation{Foundation for Research and Technology - Hellas, GR-71110 Heraklion, Greece}
\affiliation{Institute of Plasma Physics, University of Crete, GR-71003 Heraklion, Greece}

\author{Wolfram Brenig}
\email{w.brenig@tu-bs.de}
\affiliation{Institute for Theoretical Physics, Technical University Braunschweig, D-38106 Braunschweig, Germany}

\date{\today}


\begin{abstract}
We investigate the heat conductivity $\kappa$ of the Heisenberg
spin-$1/2$ ladder at finite temperature covering the {\it entire} range
of inter-chain coupling $J_\perp$, by using several numerical methods
and perturbation theory within the framework of linear response. We unveil
that a perturbative prediction $\kappa \propto J_\perp^{-2}$, based on
simple golden-rule arguments and valid in the strict limit $J_\perp \to 0$,
applies to a remarkably wide range of $J_\perp$, qualitatively and {\it
quantitatively}. In the large $J_\perp$-limit, we show power-law scaling
of opposite nature, namely, $\kappa \propto J_\perp^2$. Moreover, we
demonstrate the weak and strong coupling regimes to be connected by a
broad {\it minimum}, slightly below the isotropic point at $J_\perp =
J_\parallel$. Reducing temperature $T$, starting from $T =
\infty$, this minimum scales as $\kappa \propto T^{-2}$ down to $T$ on
the order of the exchange coupling constant. These results provide for a
comprehensive picture of $\kappa(J_\perp,T)$ of spin ladders.
\end{abstract}
\maketitle

{\it Introduction.} Thermodynamic properties of quantum many-body
systems are well understood, particularly in the vicinity of
integrable points \cite{johnston2000}. In contrast, the vast
majority of dynamical questions in these systems remain a challenge
to theoretical and experimental physics as well, in the entire range
from weak to strong breaking of integrability. These questions
consist of several timely and important issues such as eigenstate
thermalization \cite{deutsch1991, srednicki1994, rigol2008} in cold
atomic gases and, as studied in this Letter, quantum transport and
relaxation in condensed-matter materials. In this context, a
fundamental system is the one-dimensional (1D) spin-$1/2$ Heisenberg
model. It is relevant to the physics of quasi-1D quantum magnets
\cite{johnston2000}, cold atoms in optical lattices
\cite{trotzky2008}, nanostructures \cite{gambardella2006},
and to physical situations in a much broader context
\cite{kruczenski2004, kim1996}.

As typical for integrable systems, the energy current in the
spin-$1/2$ Heisenberg chain is a strictly conserved quantity
\cite{zotos1997, kluemper2002}. This implies
purely ballistic flow of heat at any temperature and provides the
theoretical basis for explaining the colossal magnetic heat
conduction observed experimentally in quasi-1D cuprates
\cite{sologubenko2000, hess2001, hess2007, hlubek2010}. In contrast
to heat flow, spin dynamics, including the existence of
ballistic \cite{shastry1990, narozhny1998, zotos1999, benz2005,
fujimoto2003, prosen2011, prosen2013, herbrych2011, karrasch2012,
karrasch2013-1, steinigeweg2014-1, steinigeweg2014-2, carmelo2014}
and diffusive transport channels \cite{sirker2009, sirker2011,
grossjohann2010, znidaric2011, steinigeweg2011, karrasch2014}, is
theoretically resolved only partially, and also under ongoing
experimental scrutiny \cite{thurber2001, maeter2012, ronzheimer2013,
hild2014, xiao2014}.

\begin{figure}[t]
\includegraphics[width=0.8\columnwidth]{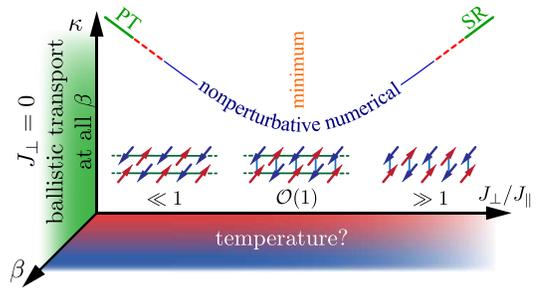}
\caption{(Color online) Thermal conductivity $\kappa$ (per chain)
versus $J_\perp/J_\parallel$ and $\beta$. PT: known perturbative
regime. Issues clarified in this Letter: extent of power-law
scaling (dashed line) close to PT and sum-rule (SR) regimes,
nonperturbative numerical treatment for entire $J_\perp$-range,
location of minimal conductivity, and temperature variation.}
\label{Fig1}
\end{figure}

Because of strict energy-current conservation in this model, the
heat conductivity $\kappa$ is highly susceptible to breaking of
integrability by, e.g., spin-phonon coupling \cite{shimshoni2003,
rozhkov2005, hlubek2012}, dimerization or disorder \cite{huang2013,
karrasch2013-2, karahalios2009}, and interactions between further
neighbors \cite{heidrichmeisner2003, steinigeweg2013}. One of the
most important perturbations is inter-chain coupling, i.e.\ $J_\perp$,
which is the key ingredient to spin-ladder compounds \cite{sologubenko2000,
hess2001}.
Since the discovery of the discontinuous transition from one
to two dimensions in quantum magnets \cite{Elbio1996}, spin ladders are a
cornerstone of correlated electron systems. They display quantum confinement
\cite{Bella2010a}, transforming gapless spinons of simple spin chains into
new massive triplons \cite{Notbohm2007,Schmidiger2013}. They provide insights
into fractionalization, quantum phase transitions \cite{Thielemann2009a},
Bose-Einstein condensation \cite{Thielemann2009b}, and disorder-induced
magnetism \cite{George1997a}. They are paradigmatic to high-T$_C$
superconductors, undoped \cite{Tranquada2004} and doped \cite{Elbio1992}.
They serve as models in other fields, e.g., cold atomic gases \cite{Garcia2004},
quantum information theory \cite{Ying2005}, and carbon nanotubes
\cite{Deshpande2009}.

Early on, perturbation theory (PT) to lowest order, i.e., a
simple golden-rule argument \cite{jung2006, jung2007}, has suggested
{\it dissipative} heat flow with a scaling $\kappa \propto J_\perp^{-2}$,
as illustrated on the l.h.s.\ of Fig.\ \ref{Fig1}. However, the relevance
of such scaling is unclear off the strict limit $J_\perp \to 0$, as is
the radius of convergence of the PT. Understanding $\kappa$ over a wider
$J_\perp$ range has been hampered by the lack of sufficiently accurate
nonperturbative methods. In particular, state-of-the-art numerical methods
have been restricted to the regime $J_\perp = {\cal O}(1)$, where
finite-size effects are moderate and spectral structures are broad, i.e.,
time scales are short \cite{zotos2004, note}. Thus, heat transport in the
transition from weakly coupled chains to strongly coupled ladders is
understood only in few and narrow regions.

In this Letter, we lift these restrictions and study the heat
conductivity $\kappa$ over the {\it entire} range of the inter-chain
coupling $J_\perp$. Using several methods within linear
response, we (a) {\it quantitatively} connect to PT in
the small-$J_\perp$ limit and (b) unveil its validity for a remarkably
wide range of $J_\perp$. In addition to the PT, scaling as $\kappa
\propto J_\perp^{-2}$, we (c) demonstrate a qualitatively different
power-law scaling $\kappa \propto J_\perp^2$ in the large $J_\perp$-limit.
Consequently, we (d) find a broad {\it minimum} of $\kappa$ in the
region $J_\perp \lesssim 1$. Reducing temperature $T$, starting
from $T = \infty$, this minimum (e) scales as $\kappa \propto T^{-2}$ down
to $T$ on the order of the exchange coupling. Thus, we provide a
comprehensive picture of $\kappa(J_\perp,T)$, beyond the
known results sketched as part of Fig.\ \ref{Fig1}.

{\it Model}. We study a Heisenberg spin-$1/2$ ladder of length
$N/2$ with periodic boundary conditions. The Hamiltonian $H =
H_\parallel + H_\perp$ consists of a leg part $H_\parallel$ and
a rung part $H_\perp$,
\begin{equation}
H_\parallel = J_\parallel \! \sum_{k=1}^{z} \sum_{i=1}^{N/2}
\mathbf{S}_{i,k} \cdot \mathbf{S}_{i+1,k} \, , \,\, H_\perp =
J_\perp \! \sum_{i=1}^{N/2} \mathbf{S}_{i,1} \cdot \mathbf{S}_{i,2}
\, ,\label{model}
\end{equation}
where $\mathbf{S}_{i,k}$ are spin-1/2 operators at site $(i,k)$,
$J_\parallel > 0$ is the antiferromagnetic leg coupling, and
$J_\perp > 0$ is the rung interaction. $z=2$ is the number of
legs. For $J_\perp = 0$, the ladder splits into integrable chains,
with a gapless ground state and spinon excitations. For $J_\parallel
= 0$, it simplifies to uncoupled dimers, with a gapped ground state
and triplon excitations. For $J_\perp, J_\parallel \neq 0$, the
ladder is nonintegrable. Generally, the model in Eq.\ (\ref{model})
preserves the total magnetization $S^z$ and is translationally
invariant. We focus on the representative sector $S^z = 0$ \cite{SM}.

The energy current has the well-known form $j = j_\parallel + j_\perp$
\cite{zotos2004},
\begin{equation}
j_\parallel = J_\parallel^2 \sum_{k=1}^z \sum_{i=1}^{N/2}
\mathbf{S}_{i-1,k} \cdot (\mathbf{S}_{i,k} \times
\mathbf{S}_{i+1,k}) \, ,
\end{equation}
\begin{equation}
j_\perp = \frac{J_\parallel J_\perp}{2} \sum_{k=1}^z
\sum_{i=1}^{N/2} (\mathbf{S}_{i-1,k}-\mathbf{S}_{i+1,k}) \cdot
(\mathbf{S}_{i,k} \times \mathbf{S}_{i,3-k})\, . \nonumber
\label{current}
\end{equation}
$j$ and $H$ commute only at the integrable point $J_\perp=0$. We
investigate the autocorrelation function at inverse temperatures
$\beta = 1/T$,
\begin{equation}
C(t) = \text{Re} \frac{\langle j(t) \, j \rangle}{N} = \text{Re}
\frac{\text{Tr} \{e^{-\beta H} j(t) \, j \}}{N \,
\text{Tr}\{e^{-\beta H}\}} \, , \label{exact}
\end{equation}
where the time argument of $j(t)$ refers to the Heisenberg picture,
$j = j(0)$, and $C(0) = 3 (J_\parallel^4 + J_\parallel^2
J_\perp^2/2)/32$ for $\beta J_\parallel \to 0$.

From $C(t)$, we first determine the Fourier transform $C(\omega)$
and then the conductivity via the low-frequency limit $\kappa/z =
\beta^2 C(\omega \to 0)$. Additionally, we can extract the
conductivity directly by $\kappa/z = \beta^2 \int_0^{t_1}\text{d}t
\, C(t)$. Here,  the cut-off time $t_1$ has to be chosen much larger
than the relaxation time $\tau$, where $C(\tau)/C(0) = 1/e$
\cite{steinigeweg2014-3}.

{\it Methods.} We calculate $C$ by complementary numerical methods,
with a particular focus on dynamical quantum typicality (DQT)
\cite{elsayed2013, steinigeweg2014-1, steinigeweg2014-2} (see also
Refs.\ \onlinecite{gemmer2003, goldstein2006, popescu2006,
reimann2007, white2009, bartsch2009, bartsch2011, sugiura2012,
hams2000}). DQT relies on the time-domain relation
\begin{equation}
C(t) = \text{Re} \frac{\langle \Phi_\beta(t) | j | \varphi_\beta(t)
\rangle}{N \, \langle \Phi_\beta(0) | \Phi_\beta(0) \rangle} +
\epsilon \, , \label{approximative}
\end{equation}
$|\Phi_\beta(t) \rangle = e^{-\imath H t -\beta H/2} \, | \psi
\rangle$, $| \varphi_\beta(t) \rangle = e^{-\imath H t} \, j \,
e^{-\beta H/2} \, |\psi \rangle$, where $|\psi \rangle$ is a {\it
single} pure state drawn at random and $\epsilon$ scales inversely
with the partition function, i.e., $\epsilon$ is exponentially small
in the number of thermally occupied eigenstates \cite{elsayed2013,
steinigeweg2014-1, steinigeweg2014-2}. The great advantage of Eq.\
(\ref{approximative}) is that it can be calculated without any
diagonalization by the use of forward-iterator algorithms. We use
a fourth-order Runge-Kutta iterator with a discrete time step
$\delta t J_\parallel = 0.01 \ll 1$. Together with sparse-matrix
representations of operators, we can reach systems sizes as large
as $N=32$. For more details on the method and its accuracy, see
Refs.\ \onlinecite{steinigeweg2014-2, SM}.

Additionally, we confirm our DQT results with numerical methods
based on Lanczos diagonalization in the frequency domain
\cite{prelovsek2013}, with the frequency resolution $\delta
\omega$ crucially depending on the number of Lanczos steps $M$,
$\delta \omega \propto 1/M$. At low $T$, we choose the finite-$T$
Lanczos method (FTLM) with $M\sim200$ \cite{SM}. At high $T$, we
also use the microcanonical Lanczos method (MCLM) with $M\sim2000$,
significantly improving $\delta \omega$.

\begin{figure}[t]
\includegraphics[width=0.85\columnwidth]{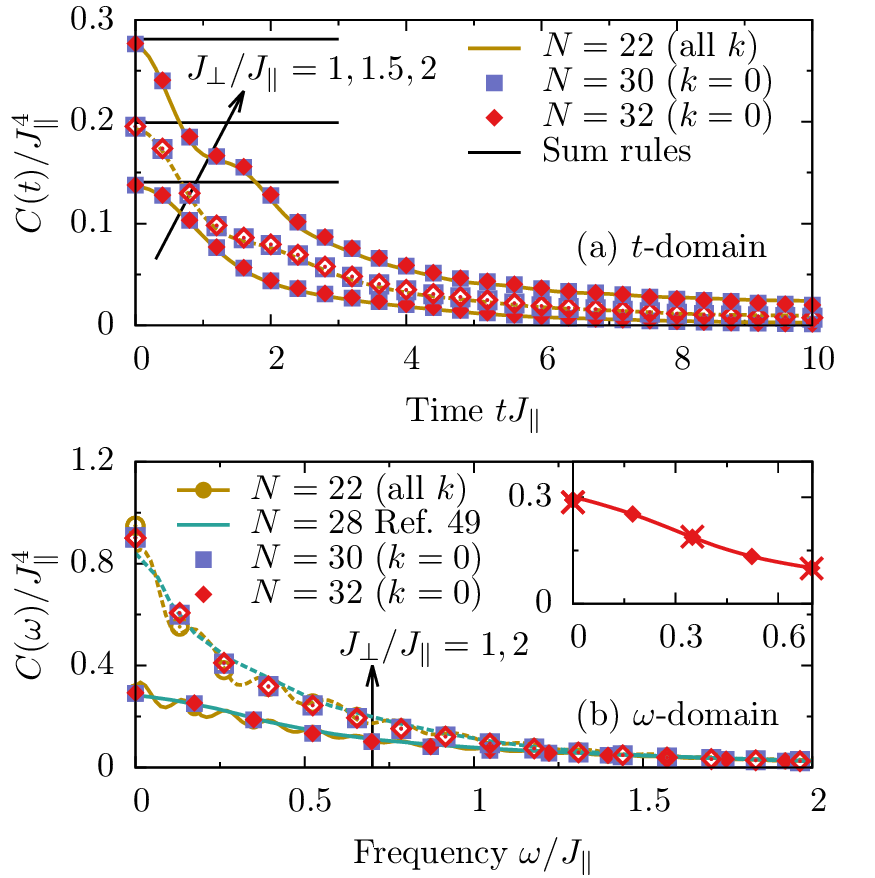}
\caption{(Color online) The (a) $t$ and (b) $\omega$ dependence
of the autocorrelation $C$ for strong $J_\perp/ J_\parallel \geq 1$,
$\beta J_\parallel \to 0$, and $N \leq 32$, as obtained from DQT.
Spectra in (b) are obtained by Fourier transforming finite-$t$
data $t \leq 10\tau \sim 20/J_\parallel$ (symbols); Inset:
Low-$\omega$ limit for $J_\perp/J_\parallel = 1$, the largest $N=32$,
and $t \leq 5\tau$ (crosses), $10 \tau$ (other symbols), $50 \tau$
(curves); Main panel: Spectra from Lanczos methods ($N=28$ MCLM of
Ref.\ \onlinecite{zotos2004}, $N=22$ FTLM) are shown (curves). Note that
method-related errors are negligibly small \cite{SM}.} \label{Fig2}
\end{figure}

{\it Results.} We begin with $J_\perp/J_\parallel \geq 1$ and
$\beta J_\parallel \to 0$. In Fig.\ \ref{Fig2} (a) we summarize
our DQT results on $C(t)$ for different $J_\perp/J_\parallel =
1$, $1.5$, $2$. Several comments are in order. First, the
initial value $C(0)$ agrees with the high-$T$ sum rule and
therefore increases with $J_\perp$. Second, all $C(t)$ depicted
decay to zero on a time scale $5 \tau \sim 10/J_\parallel$. Third,
the $C(t)$ curves do not change when the number of sites is
increased from $N=22$ to $32$. Thus, we observe very little
finite-size effects, i.e., we can safely consider our results as
results on $C(t)$ for $N \to \infty$. Note that for $N \geq 30$
we consider a single translation subspace $k$ since, for these
$N$, $C(t)$ is $k$ independent at $\beta \to 0$
\cite{steinigeweg2014-1, steinigeweg2014-2}.

Next, we discuss the spectrum $C(\omega)$. To this end, we show in
Fig.\ \ref{Fig2} (b) for $J_\perp/J_\parallel = 1$, $2$ the Fourier
transform of our DQT data for times $t \leq 10 \tau \sim 20/J_\parallel$.
These times correspond to a frequency resolution $\delta \omega \sim
0.15 J_\parallel$. For this resolution, the Fourier transform
is a smooth function of $\omega$ and displays a well-behaved limit for
$\omega \to 0$, i.e., $C(\omega \to 0) = C(\omega = 0)$. Moreover, this
limit and $C(\omega)$ in general do not depend on system size for $N
\geq 22$. The inset of Fig.\ \ref{Fig2} (b) clarifies the impact of the
$\omega$ resolution by displaying additional Fourier transforms of DQT
data, evaluated for shorter ($t \leq 5\tau$) and longer ($t \leq 50
\tau$) times at $J_\perp/J_\parallel = 1$ and for the largest $N = 32$.
Clearly, the low-$\omega$ limit is independent of the $\omega$ resolution
resulting from the specific choice of $t$. This robustness, together
with the $N$ independence, allows us to reliably extract a quantitative
value for the dc conductivity at $J_\perp/J_\parallel =1$, $\kappa/z
\beta^2 J_\parallel^3 = 0.29$.

To additionally demonstrate the validity of our DQT approach, we
compare to our FTLM results and to existing MCLM spectra from the
literature \cite{zotos2004} in Fig.\ \ref{Fig2} (b). Obviously,
the agreement is very good.

\begin{figure}[t]
\includegraphics[width=0.85\columnwidth]{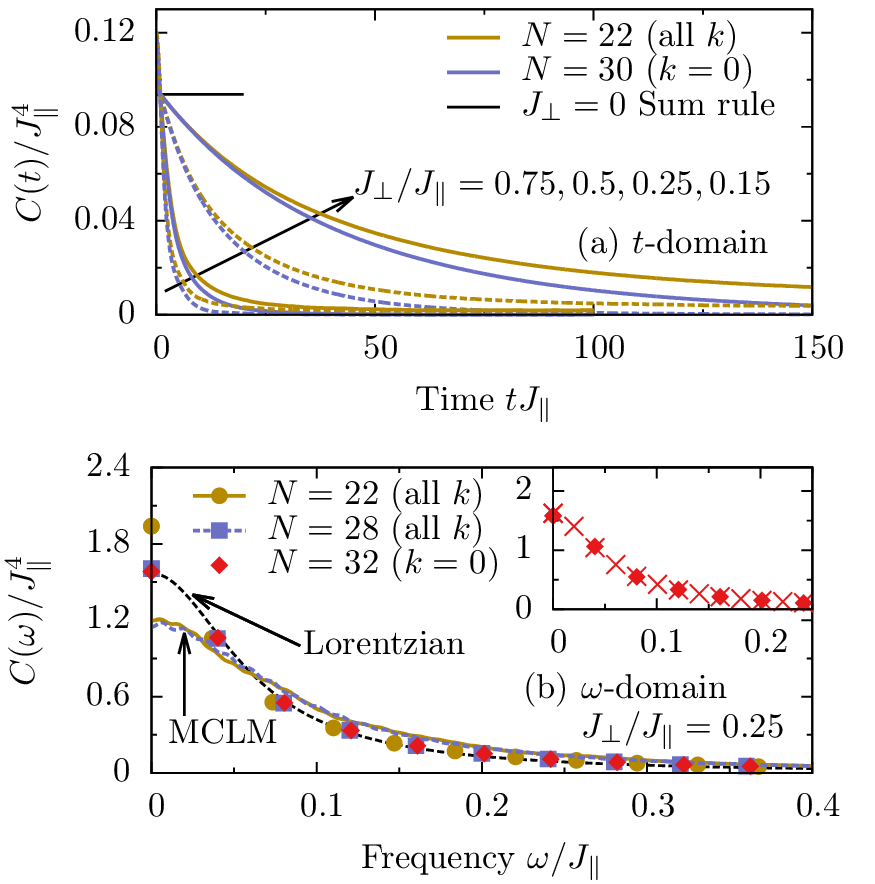}
\caption{(Color online) (a) The $t$ dependence of $C$ for various
small $J_\perp/ J_\parallel = 0.15$, \ldots, $0.75$, obtained from
DQT for $\beta J_\parallel \to 0$ and $N \leq 30$. (b) Spectrum for
$J_\perp/ J_\parallel = 0.25$, obtained by Fourier transforming
finite-$t$ data $t \leq 5 \tau \sim 80/ J_\parallel$ (symbols);
Inset: Low-$\omega$ limit for the largest $N=32$ and $t \leq 5 \tau$
(diamonds), $10 \tau$ (crosses); Main panel: Spectrum from $N=22$
and $28$ MCLM and a Lorentzian fit are shown (curves).} \label{Fig3}
\end{figure}

Now, we turn to small $J_\perp/J_\parallel < 1$. In Fig.\
\ref{Fig3} (a) we depict our DQT results on $C(t)$ for various $J_\perp
/J_\parallel = 0.15, \ldots, 0.75$. The initial value $C(0)$ approaches
the $J_\perp = 0$ sum rule when $J_\perp$ is reduced. Furthermore, the
decay is slower for smaller $J_\perp$ and finite-size effects are
naturally stronger in the vicinity of the integrable point $J_\perp =
0$. For the smallest $J_\perp/J_\parallel=0.15$ depicted, these
finite-size effects are still moderate when comparing $C(t)$ for
$N=22, 30$. In Fig.\ \ref{Fig3} (b) we show the Fourier
transform of $C(t \leq 5 \tau \sim 80 J_\parallel)$ for $J_\perp/
J_\parallel = 0.25$. For the largest $N=32$, this Fourier transform is
well described by a Lorentzian line shape and, again, the low-$\omega$
limit does not depend on $t$. Since $C(\omega)$ has
a narrow spectrum, MCLM with a high $\omega$
resolution ($M = 2000$) is a better choice for comparison than FTLM
($M = 200$) \cite{SM}, and agrees well with DQT.
Note that resolving narrow spectral features by DQT is a new
concept of our Letter, which can be applied in a much broader context.

\begin{figure}[t]
\includegraphics[width=0.85\columnwidth]{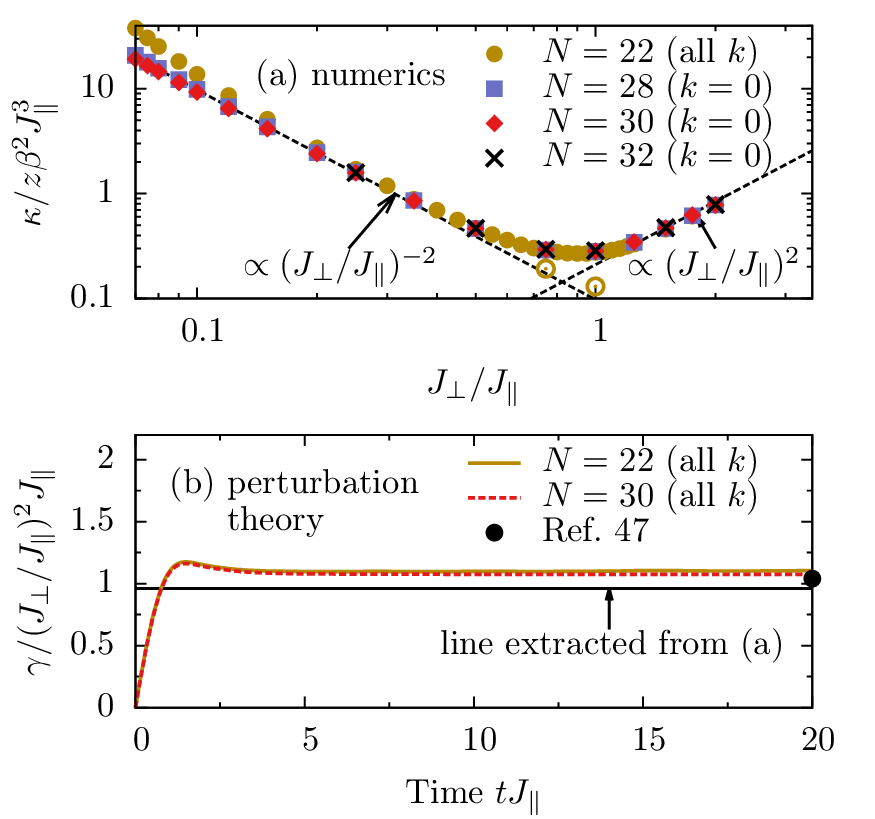}
\caption{(Color online) (a) Scaling of the conductivity $\kappa$
with $J_\perp$, obtained from DQT and finite-$t$ data $t \leq 5
\tau$ for $\beta J_\parallel \to 0$ and $N \leq 32$ (closed symbols).
Results for the simplified operator $j' = j_\parallel$ are also
depicted at $J_\perp/J_\parallel \sim 1$ (open symbols). Additionally,
power laws $0.097 (J_\perp/J_\parallel)^{-2}$ and $0.21 (J_\perp/
J_\parallel)^2$ are shown (lines). (b) PT for the scattering rate
$\gamma$, carried out using DQT. The PT of Ref.\ \onlinecite{jung2006}
is also depicted (bullet).} \label{Fig4}
\end{figure}

Next, we discuss the scaling of the conductivity $\kappa$ over the
{\it entire} range of $J_\perp$. In Fig.\ \ref{Fig4} (a) we summarize
$\kappa(J_\perp)$ as inferred from DQT data for $C(t \leq 5 \tau)$. Here,
we observe a broad minimum of $\kappa(J_\perp)$, centered between two
regimes with power-law scaling at large and small $J_\perp$. The scaling
$\propto J_\perp^2$ in the large $J_\perp$ limit is a direct consequence
of the static sum rule $C(0) \propto J_\perp^2$, noted following
Eq.\ (\ref{exact}). The scaling $\propto J_\perp^{-2}$ for small $J_\perp$,
however, is not simply related to $C(0)$ since $C(0) \approx
\text{const.}$  for such $J_\perp$. Particularly, we find this scaling to
hold over a remarkably wide range of $0.07 \leq J_\perp/J_\parallel \lesssim
0.35$. This finding is a central result of this Letter. Below $J_\perp/
J_\parallel < 0.07$, computational efforts for $5 \tau$ data are very high
and finite-size effects are too large, even for N accessible to
DQT.

To gain further insight into the scaling at small $J_\perp$, we
calculate the scattering rate $\gamma = 1/\tau$ to lowest order in
$J_\perp$, i.e., $J_\perp^2$, following the PTs in Refs.\
\onlinecite{jung2006, jung2007, steinigeweg2010, steinigeweg2011-1}.
This rate reads ($\beta J_\parallel \to 0$)
\begin{equation}
\gamma = \lim_{t_1 \to \infty} \int_0^{t_1} \! \text{d}t_\parallel
\frac{\text{Tr} \{ \imath [j_\parallel, H_\perp](t_\parallel) \,
\imath [j_\parallel, H_\perp] \} }{\text{Tr} \{ j_\parallel^2 \}}
\propto J_\perp^2 \, ,
\label{pt}
\end{equation}
where $t_\parallel$ refers to the Heisenberg picture of $H_\parallel$.
Figure \ref{Fig4} (b) shows $\gamma$ evaluated by DQT
applied to Eq.\ (\ref{pt}) for large $N \leq 30$. Note that
this application of DQT is a new concept of our Letter \cite{SM}. As
shown in Fig.\ \ref{Fig4}, we find good agreement with previous evaluation
of $\gamma$ in Ref.\ \onlinecite{jung2006} based on smaller systems. Most
notably, however, $\gamma$ well agrees with the scattering rate $\gamma'$
as extracted directly from $\kappa$ in Fig.\ \ref{Fig4} (a) via the
relation $\gamma' = z \beta^2 C(0)/\kappa$. This agreement is another
main result of our Letter. Note that PT holds up to $J_\perp/ J_\parallel
\sim 1$ for the simplified current $j = j_\parallel$, see Fig.\ \ref{Fig4}
(a), which is the regime where the system behaves Markovian, i.e.,
has no memory. For the explicit calculation of the memory kernel, see
Ref.\ \onlinecite{SM}.

\begin{figure}[t]
\includegraphics[width=0.85\columnwidth]{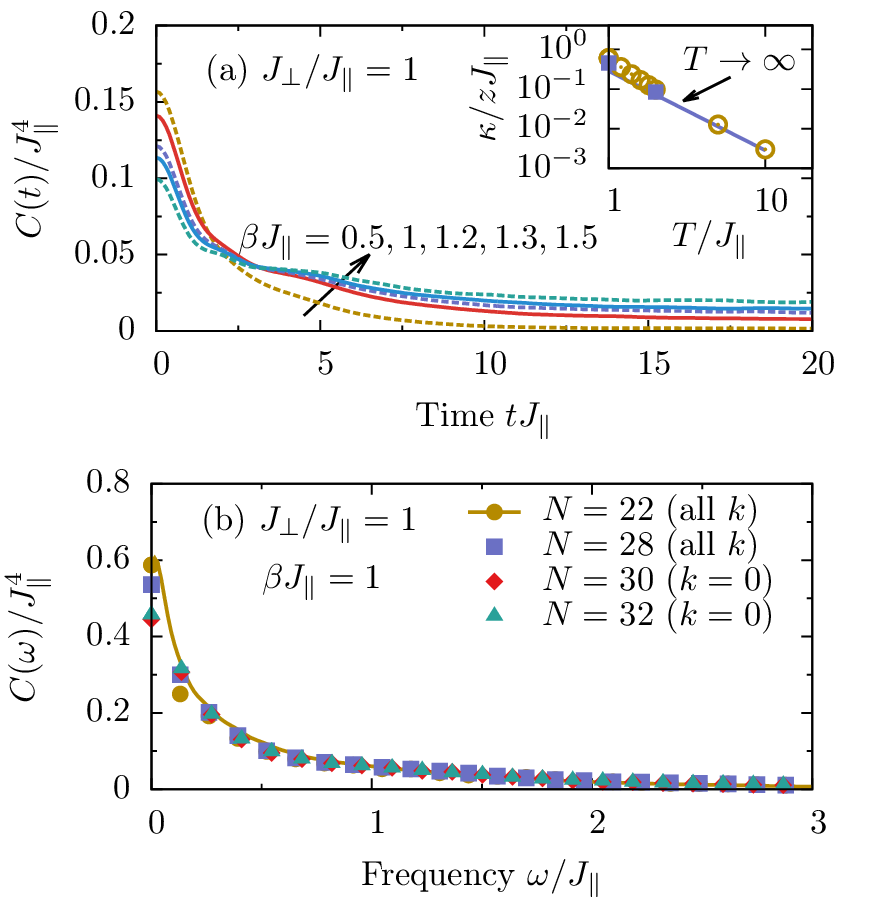}
\caption{(Color online) (a) The $t$ dependence of $C$ for $\beta
J_\parallel = 0.5, \ldots, 1.5$, obtained from DQT for $J_\perp/
J_\parallel = 1$ and $N=28$. (b) Spectrum for $\beta J_\parallel = 1$,
obtained by Fourier transforming finite-$t$ data $t \leq 5 \tau
\sim 12/J_\parallel$ for $N \leq 32$. Additionally, a spectrum from
$N=22$ FTLM is depicted (curve). Inset: $T$ dependence of the conductivity
$\kappa$, calculated by $N=32$ DQT (closed symbols, curve) and $N=22$ FTLM
(open symbols).}
\label{Fig5}
\end{figure}

Now we turn to $\beta J_\parallel \neq 0$, focusing on
$J_\perp/J_\parallel = 1$. In Fig.\ \ref{Fig5} (a) we depict
our DQT results for $C(t)$ for $\beta J_\parallel = 0.5, \ldots, 1.5$.
While $C(0)$ decreases as $\beta$ is increased, the relaxation time
shows the tendency to increase with $\beta$. However, significant
finite-size effects appear as nondecaying Drude weights. Since these
Drude weights exceed $20\%$ of $C(0)$ at $\beta J_\parallel \sim 1.5$,
we restrict ourselves to $\beta J_\parallel \leq 1$. For such $\beta$,
once again, FTLM agrees with the Fourier transform of our DQT data,
which also shows a $N$ independent dc limit for large $N \sim 30$,
see Fig.\ \ref{Fig5} (b). Finally, in the inset of Fig.\ \ref{Fig5}
(a) we show the $T$ dependence of the conductivity $\kappa$. Remarkably,
in the $T$ range accessible to our methods, we observe no significant
deviations from the high-$T$ behavior $\kappa \propto \beta^2$. While
these $T$ are low from a numerical point of view, they are still too
high for a comparison to experiments on yet available materials, where
the exchange coupling constant is large.

{\it Conclusion.} We studied the heat conductivity $\kappa$ of
the Heisenberg spin-$1/2$ ladder at finite temperature and over the
{\it entire} range of the rung interaction $J_\perp$, using several
methods within linear response. We detailed the power-law scalings
$\kappa \propto J_\perp^{-2}$ and $\kappa \propto J_\perp^{2}$ at
weak and strong $J_\perp$, respectively. We found a broad {\it
minimum} of $\kappa$ in the region $J_\perp \sim 1$, with a scaling
of its temperature dependence as $\kappa \propto T^{-2}$ down to
$T$ on the order of the exchange coupling. Thus, we provided a
comprehensive picture of $\kappa(J_\perp,T)$.

{\it Acknowledgments.}  
W.\ Brenig acknowledges support by the DFG through SFB 1143, the Lower
Saxony PhD program SCNS, and the Platform for Superconductivity and
Magnetism, Dresden.

X.\ Zotos acknowledges support by the Greek national funds through
the Operational Program ``Education and Lifelong Learning'' of the
NSRF under ``Funding of proposals that have received a positive
evaluation in the 3rd and 4th call of ERC Grant Schemes''; and
together with J.\ Herbrych support by the EU program
FP7-REGPOT-2012-2013-1 under Grant No.\ 316165.

\newpage

\setcounter{figure}{0}
\setcounter{equation}{0}
\renewcommand*{\citenumfont}[1]{S#1}
\renewcommand*{\bibnumfmt}[1]{[S#1]}
\renewcommand{\thefigure}{S\arabic{figure}}
\renewcommand{\theequation}{S\arabic{equation}}

\section{\large Supplemental Material}

\section{Perturbation Theory}

\subsection{Leading-Order Scattering Rate in the Markov Limit}

In this section we discuss the perturbation theory for the energy current $j =
j_\parallel + j_\perp$ in detail. In the limit of small inter-chain coupling,
$J_\perp \to 0$, the leg part $j_\parallel$ is the dominant contribution, i.e.,
\begin{equation}
j = j_\parallel + {\cal O}(J_\perp) \, .
\end{equation}
Because the leg part $j_\parallel$ is strictly conserved for the leg
Hamiltonian $H_\parallel$, $[j_\parallel, H_\parallel] = 0$, the rung
Hamiltonian $H_\perp$ is the only origin of the scattering of $j_\parallel$.
This scattering can be treated perturbatively if the inter-chain coupling
$J_\perp$ is a sufficiently small parameter. In the time domain, we can
formulate such a perturbation theory in terms of the integro-differential
equation
\begin{equation}
\dot{C}(t) = - \int \limits_0^t \! \text{d}t' \, K(t-t') \,
C(t) \label{master1}
\end{equation}
for the autocorrelation function $C(t) = \text{Re} \langle j_\parallel(t)
j_\parallel \rangle / N$ of the leg part $j_\parallel$, where the memory kernel
$K(t)$ occurs in the time convolution on the right side. To leading order of the
perturbation $J_\perp$, $J_\perp^2$, and in the high-temperature limit, $\beta
\to 0$, this memory kernel reads 
\cite{Ssteinigeweg2010, Ssteinigeweg2011-1}
\begin{equation}
K(t) = \frac{\text{Tr} \{ \imath [j_\parallel, H_\perp](t_\parallel) \, \imath
[j_\parallel, H_\perp] \} }{\text{Tr} \{ j_\parallel^2 \}} \propto J_\perp^2
\, , \label{memory}
\end{equation}
where $t_\parallel$ indicates the Heisenberg picture with respect to
$H_\parallel$, i.e.,
\begin{equation}
\imath [j_\parallel, H_\perp](t_\parallel) = e^{\imath H_\parallel t} \, \imath
[j_\parallel, H_\perp] \, e^{-\imath H_\parallel t} \, . \label{Heisenberg}
\end{equation}
Assuming that $K(t)$ fully decays on a finite time scale $\tau_K$, i.e., using
the Markov approximation, Eq.\ (\ref{master1}) simplifies for small $J_\perp$,
where $C(t)$ decays on a very long time scale $\tau \gg \tau_K$. Thus, Eq.\
(\ref{master1}) factorizes into
\begin{equation}
\dot{C}(t) = -\gamma \, C(t) \, , \label{master2}
\end{equation}
where $\gamma$ is the scattering rate
\begin{equation}
\gamma = \frac{1}{\tau} = \lim_{t \to \infty} \int_0^t \text{d}t' \, K(t')
\propto J_\perp^2 \, ,
\end{equation}
cf.\ Eq. (5) of our Letter. Obviously, Eq.\ (\ref{master2}) implies the
exponential relaxation
\begin{equation}
C(t) = C(0) \, e^{-\gamma t} \, .
\end{equation}
Consequently, the heat conductivity becomes
\begin{equation}
\frac{\kappa}{z \, \beta^2} = \lim_{t \to \infty} \int_0^t \text{d}t' \, C(t')
= \frac{C(0)}{\gamma} \propto \frac{1}{J_\perp^2} \, .
\end{equation}
Note that, in the Markov approximation, the qualitative scaling $\propto 1/
J_\perp^2$ does not depend on details of the memory kernel $K(t)$ while the
quantitative value of the scattering rate $\gamma$ clearly does.

\subsection{Strong Perturbations}

For strong inter-chain coupling $J_\perp$, the perturbation theory discussed
above necessarily breaks down for two different reasons: (i) The memory kernel 
$K(t)$ in Eq.\ (\ref{memory}) does not incorporate higher-order contributions.
(ii) The current operator $j$ is approximated by the leg part $j_\parallel$.
In fact,
\begin{equation}
j \approx j_\perp
\end{equation}
in the limit of large $J_\perp$. In this limit, $C(0) = \text{const.}$ turns
into $C(0) \propto J_\perp^2$. This scaling with $J_\perp$ reflects that the
dominant energy contribution is the bond energy in the rungs.

\subsection{Dynamical Quantum Typicality}

Verifying the Markov approximation and determining the quantitative value of
the scattering rate $\gamma$ necessarily requires full knowledge about the time
dependence of the memory kernel $K(t)$. Even though this time dependence is
generated by the integrable Hamiltonian $H_\parallel$, cf.\ Eqs.\
(\ref{memory}) and (\ref{Heisenberg}), an exact calculation of $K(t)$ is
unfeasible due to the complexity of $H_\parallel$. Therefore, in praxis, $K(t)$
has to be calculated numerically. The standard approach is the exact
diagonalization of $H_\parallel$ \cite{Sjung2006, Sjung2007}. However, since
$H_\parallel$ is a many-body Hamiltonian, this approach is only feasible for at
most $N \sim 16$ sites and finite-size effects can be large for such $N$.

To overcome the limitation of exact diagonalization to small $N$, we first need
to note that the memory kernel $K(t)$ for $\beta \to 0$ is just the autocorrelation
function of the Hermitian operator
\begin{equation}
j' = \frac{\imath [j_\parallel, H_\perp]}{\sqrt{\text{Tr} \{ j_\parallel^2 \}}}
\, , 
\end{equation}
i.e., K(t) = $\text{Tr} \{j'(t_\parallel) j' \}$. Hence, remarkably, we can
use the concept of dynamical quantum typicality to calculate $K(t)$.
Specifically, in analogy to Eq.\ (4) of our Letter, we get the relation
for $\beta \to 0$
\begin{equation}
K(t) = \frac{\langle \Phi(t) | j' | \varphi(t) \rangle}{\langle \Phi(0) |
\Phi(0) \rangle} + \epsilon \label{K1}
\end{equation}
with the two auxiliary states
\begin{eqnarray}
|\Phi(t) \rangle &=& e^{-\imath H_\parallel t} \, | \psi \rangle \, ,
\label{K2} \\
| \varphi(t) \rangle &=& e^{-\imath H_\parallel t} \, j' \, |\psi
\rangle \, , \label{K3}
\end{eqnarray}
where $|\psi \rangle$ is a {\it single} pure state drawn at random. Again,
$\epsilon$ scales inversely with the partition function, i.e., $\epsilon$
is exponentially small in the number of thermally occupied eigenstates
\cite{Selsayed2013, Ssteinigeweg2014-1, Ssteinigeweg2014-2}.

\begin{figure}[t]
\includegraphics[width=0.9\columnwidth]{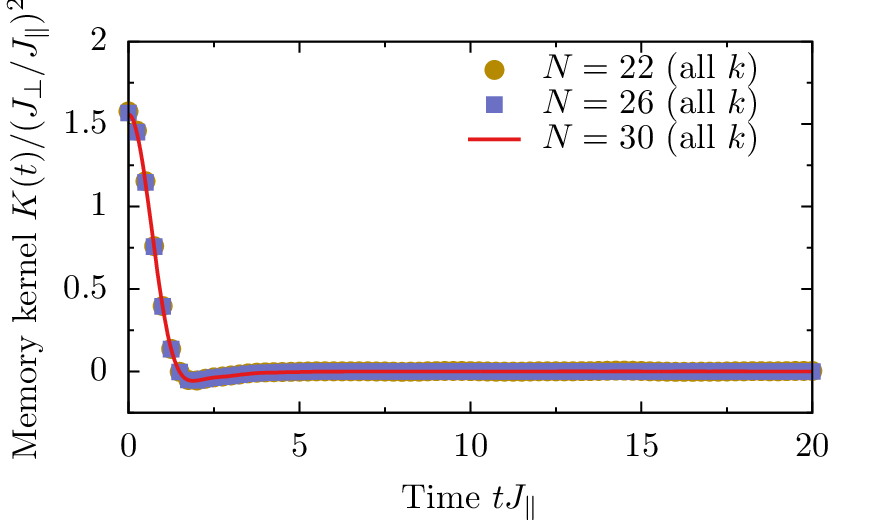}
\caption{(Color online) Full time dependence of the memory kernel $K(t)$,
according to the leading-order prediction in Eq.\ (\ref{memory}), for high
temperatures $\beta \to 0$ and different lattice sites $N=22$, $26$, $30$.
The data depicted is numerically calculated using the typicality relation in
Eqs.\ (\ref{K1}), (\ref{K2}), (\ref{K3}) and a fourth-order Runge-Kutta
iterator with a discrete time step $\delta t \, J_\parallel = 0.01 \ll 1$.
Apparently, $K(t)$ does not depend on $N$ and fully decays on a rather short
time scale $\tau_K \, J_\parallel \sim 2$. As compared to exact diagonalization
of, say, $N = 16$ sites, the Hilbert-space dimension accessible is larger by a
factor of ca.\ $16,000$.}
\label{Fig6}
\end{figure}

The typicality relation in Eqs.\ (\ref{K1}), (\ref{K2}), and (\ref{K3})
can be calculated for as many sites as $N=30$, using a fourth-order
Runge-Kutta iterator with a discrete time step $\delta t J_\parallel =
0.01 \ll 1$ and sparse-matrix representations of the operators $H_\parallel$
and $j'$ \cite{Ssteinigeweg2014-2}. In Fig.\ \ref{Fig6} we depict our results
on the time dependence of $K(t)$ for $\beta \to 0$ and different $N$.
Apparently, $K(t)$ does not depend on $N$ and fully decays on a rather short
time scale $\tau_K \, J_\parallel \sim 2$. Thus, the Markov approximation is
indeed justified. Note that the area under the $K(t)$ curve is the scattering
rate $\gamma$ shown in Fig.\ 4 (b) of our Letter.

\section{Error Analysis}

\subsection{Specific Realization of the Initial State}

\begin{figure}[t]
\includegraphics[width=0.9\columnwidth]{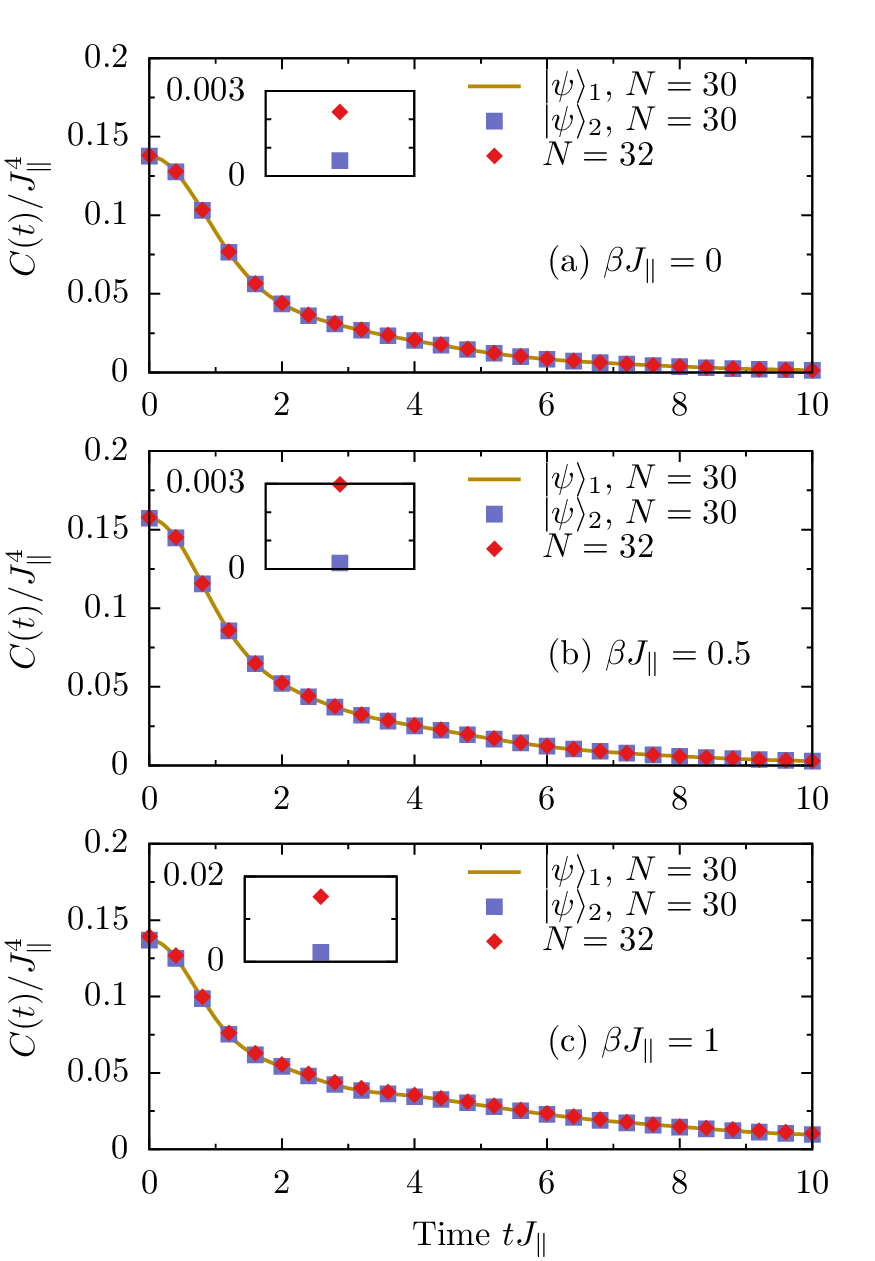}
\caption{(Color online) Error analysis: The time-dependent energy-current
autocorrelation function $C(t)$ is numerically calculated according to the
typicality relation in Eq.\ (4) of our Letter for two random initial states
$| \psi \rangle_1 \neq | \psi \rangle_2$, $N = 30$ lattice sites, equal
exchange couplings $J_\perp = J_\parallel$, and different temperatures (a)
$\beta \, J_\parallel = 0$, (b) $\beta \, J_\parallel = 0.5$, (c) $\beta \,
J_\parallel = 1$. In addition, results for $N=32$ and a single initial state
are shown. All results in (a)-(c) correspond to the sector $S^z = 0$ and
$k = 0$. Insets: Relative deviation of the initial value $C(0)$, as obtained
from $| \psi \rangle_2$ or $N=32$, from the $C(0)$ of $| \psi \rangle_1$.
(Note that the $x$ axis is meaningless.) For the $\beta$ range depicted,
the initial-state dependence is much below $1\%$ and smaller than the, also
small, finite-size effects.}
\label{Fig7}
\end{figure}

Our main numerical method used is essentially based on the
typicality relation in Eq.\ (4) of our Letter, where the random error
$\epsilon$ occurs. In this section we discuss this error in detail. The
probability distribution $p(\epsilon)$, i.e., the probability to get
an error of size $\epsilon$, has the mean
\begin{equation}
\bar{\epsilon} = 0 \, .
\end{equation}
Hence, if averaging is performed over sufficiently many random initial
states $| \psi \rangle_i$, then any error vanishes. Note that we do
not need to perform such averaging for reasons outlined below. The width
of the probability distribution $p(\epsilon)$ is {\it upper} bounded
by \cite{Selsayed2013, Ssteinigeweg2014-1, Ssteinigeweg2014-2}
\begin{equation}
\sigma(\epsilon) \leq {\cal O} \left ( \frac{\sqrt{\text{Re} \, \langle
j(t) \, j \, j(t) \, j \rangle}} {N \, \sqrt{d_\text{eff}}}
\right ) \, , \label{error}
\end{equation}
where the effective dimension
\begin{equation}
d_\text{eff} = \text{Tr} \{ e^{-\beta (H - E_0)} \}
\label{deff}
\end{equation}
is the partition function and $E_0$ denotes the ground-state energy. Hence,
the maximum error $\sigma(\epsilon)$ decreases {\it faster} with system
size than $1/\sqrt{d_\text{eff}}$ does. In the limit of high temperatures,
$\beta \to 0$, $d_\text{eff} = 2^N$ is a huge number for $N \sim 30$ and
the maximum error consequently is tiny. In fact, it has been shown in Ref.\
\onlinecite{Ssteinigeweg2014-2} that significant errors only occur for 
effective dimensions below $d_\text{eff} \sim 10,000$, e.g., for rather
small $N$. However, for large $N$, $d_\text{eff}$ can also become small
for two reasons relevant to the study in our Letter.

(i) To reduce computational effort for large $N \geq 30$, we restrict our
investigation to a single but {\it representative} symmetry subspace, i.e.,
to the quantum numbers $S^z = 0$ and $k = 0$. While this restriction does
not have impact on the exact autocorrelation function $C(t)$,
the dimension of the subspace
\begin{equation}
d_{0,0} \approx \frac{1}{N/2} \binom{N}{N/2} \ll 2^N
\end{equation}
is much smaller than the full Hilbert-space dimension. In the
high-temperature limit, $\beta \to 0$, $d_\text{eff} = d_{0,0}$ is still
a large number for $N \sim 30$.

(ii) If temperature is reduced from infinity at fixed $N$, $d_\text{eff}$
gradually deceases and eventually becomes $1$ at zero temperature. Therefore,
$d_\text{eff}$ becomes a small number for sufficiently low temperatures
and, as a consequence, the upper bound in Eq.\ (\ref{error}) does not imply
a small $\epsilon$ any further. Note that $\epsilon$ can still be small
since $\epsilon = 0$ at zero temperature.

Due to (i) and (ii), it is important to verify in praxis that $\epsilon$
is indeed a negligibly small error. This verification is most conveniently
done by repeating the calculation of the autocorrelation function $C(t)$
for a second random initial state $| \psi \rangle_2 \neq | \psi \rangle_1$.
In Fig.\ \ref{Fig7} we depict $C(t)$, as obtained from $|\psi \rangle_1$
and $| \psi \rangle_2$, for $N = 30$ lattice sites, quantum numbers $S^z
= 0$ and $k = 0$, equal couplings $J_\perp = J_\parallel$, and different
temperatures $\beta \, J_\parallel \leq 1$.  For this temperature range,
we extract the heat conductivity in our Letter. Clearly, the $C(t)$ curves
for $|\psi \rangle_1$ and $|\psi \rangle_2$ in Fig.\ \ref{Fig7} are very
close to each other. This independence of the specific realization of
the random initial state proves a small $\epsilon$ for temperatures $\beta
\, J_\parallel = 1$.

It is also instructive to quantify the size of errors. To this end, let
us consider the relative error of the initial value given by
\begin{equation}
\epsilon_r(0) = \frac{| C(0,| \psi \rangle_2) - C(0,| \psi \rangle_1) |
}{C(0,| \psi \rangle_1)} \, .
\end{equation}
As shown in the insets of Fig.\ \ref{Fig7}, $\epsilon_r(0)$ is much
smaller than $1\%$ and particularly smaller than the, also small,
finite-size effects.

\begin{figure}[b]
\includegraphics[width=0.9\columnwidth]{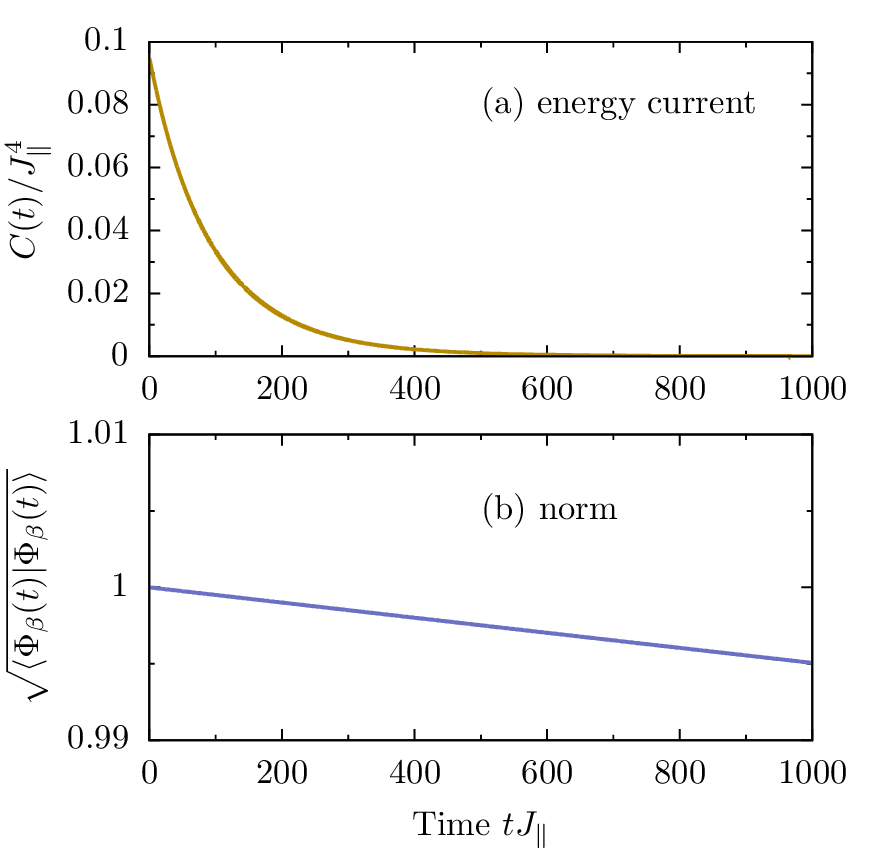}
\caption{(Color online) Error analysis: Real-time dependence of (a)
energy-current autocorrelation function $C(t)$ and (b) norm $\sqrt{\langle
\Phi_\beta(t) | \Phi_\beta(t) \rangle}$ of the pure state $|\Phi_\beta(t)
\rangle$ in Eq. (4) of our Letter for small inter-chain coupling $J_\perp
/ J_\parallel = 0.1$, $N=30$ lattice sites, sector $S^z = 0$ and $k = 0$,
and high temperatures $\beta \to 0$. Although $C(t)$ decays on a long
time scale, the norm $\sqrt{\langle \Phi_\beta(t) | \Phi_\beta(t) \rangle}$
does not deviate more than $0.25 \%$ from $1$ on this time scale. Hence,
our choice of the Runge-Kutta time step $\delta t \, J_\parallel = 0.01
\ll 1$ is reasonable.}
\label{Fig8}
\end{figure}

\subsection{Runge-Kutta Time Step}

The typicality relation in Eq.\ (4) of our Letter requires to propagate
pure states in real and imaginary time. We perform the propagation by a
fourth-order Runge-Kutta iterator with a discrete time step $\delta t \,
J_\parallel = 0.01 \ll 1$. This time step is a potential source for errors
if the relaxation time of the energy current is very long, i.e., for very
small inter-chain couplings $J_\perp / J_\parallel \ll 1$. To ensure
sufficiently high accuracy, we verified for all $J_\perp$ that the norm of
the two pure states $| \Phi_\beta(t) \rangle$ and $| \phi_\beta(t) \rangle$
propagated does not deviate significantly from $1$ on times up to the
relaxation time of the energy current. In Fig.\ \ref{Fig7} we illustrate
that, even for the demanding case $J_\perp / J_\parallel = 0.1$, we find
that deviations from $1$ are less than $1 \%$. Note that reducing the
time step for the large $N=30$ depicted is unfeasible for $J_\perp /
J_\parallel \sim 0.1$.

\begin{figure}[t]
\includegraphics[width=0.9\columnwidth]{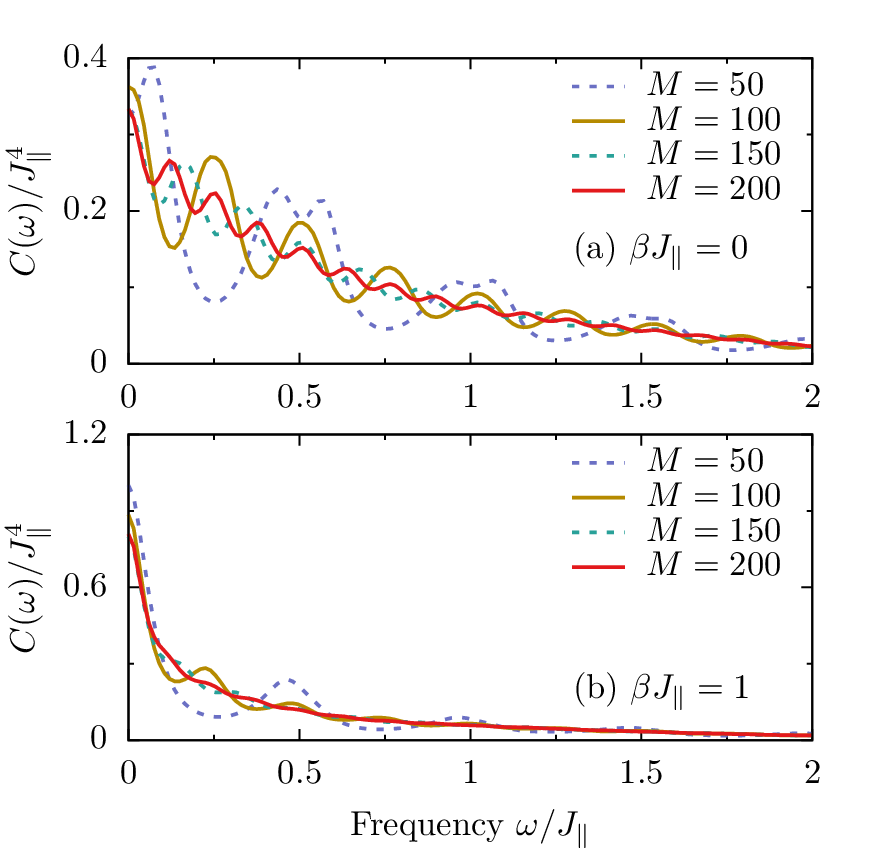}
\caption{(Color online) Error analysis: FTLM results for the frequency-dependent
energy-current autocorrelation function $C(\omega)$ for different Lanczos steps
$M = 50$, $100$, $150$, $200$ and two temperatures (a) $\beta \, J_\parallel = 0$
and (b) $\beta \, J_\parallel = 1$. All results depicted correspond to strong
coupling $J_\perp / J_\parallel = 1$, $N = 22$ lattice sites, and the sector
$S^z = 0$ (all $k$).}
\label{Fig9}
\end{figure}

\begin{figure}[b]
\includegraphics[width=0.9\columnwidth]{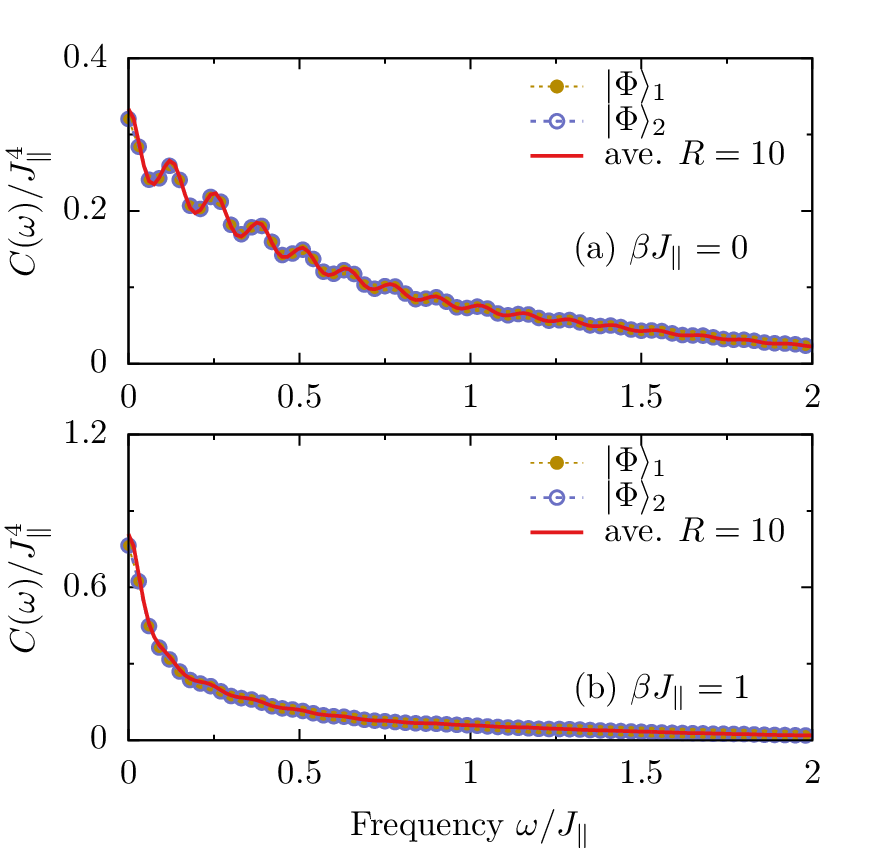}
\caption{(Color online) Error analysis: FTLM results for the frequency-dependent
energy-current autocorrelation function $C(\omega)$ for two different random
states $| \Phi \rangle_1 \neq | \Phi \rangle_2$ ($R = 1$), an average over several 
$| \Phi \rangle_i$ ($R = 10$), and two temperatures (a) $\beta \, J_\parallel = 0$
and (b) $\beta \, J_\parallel = 1$.  All results depicted correspond to strong
coupling $J_\perp / J_\parallel = 1$, $N = 22$ lattice sites, and the sector
$S^z = 0$ (all $k$).}
\label{Fig10}
\end{figure}

\subsection{Lanczos-Related Errors}
Within the Lanczos-diagonalization techniques used in our Letter, the
origin of potential errors is twofold and related to \cite{Sprelovsek2013}:
(i) spectral resolution $\delta \omega$ and (ii) number of effective terms
in the thermodynamic sum $Z^*$.

(i) The spectral resolution is given by 
\begin{equation}
\delta \omega = \frac{\Delta E}{M}\,,
\end{equation}
where $M$ is the number of Lanczos steps used. Note that we use $M = 200$
for FTLM and $M = 2000$ for MCLM in our Letter. $\Delta E = E_{\text{max}}
- E_{\text{min}}$ is the full energy span of the Hamiltonian, i.e.,
$E_\text{min}$ and $E_\text{max}$ are the smallest and largest eigenenergy,
respectively. This span depends only weakly on the inter-chain coupling
$J_\perp$, e.g., for $N=22$ we find $\Delta E/ J_\parallel \approx 30$ for
the ladder case $J_\perp/J_\parallel = 1$ and $\Delta E / J_\parallel
\approx 20$ for the chain case $J_\perp/J_\parallel = 0$. This span,
together with $M = 200$, yield $\delta \omega / J_\parallel \approx 0.15$.
It is evident from Fig.\ 2 of our Letter that this spectral resolution is
sufficient for large $J_\perp/J_\parallel = {\cal O}(1)$. However, the
spectrum $C(\omega)$ is narrow for small $J_\perp / J_\parallel = 0.25$ and
has a width of roughly $0.2 \, J_\parallel$. Thus, we have to turn to MCLM
and $M=2000$ for sufficiently high spectral resolution.

In Fig.\ \ref{Fig9} we show the $M$ dependence of the FTLM result for
strong coupling $J_\perp / J_\parallel = 1$, two temperatures $\beta \,
J_\parallel = 0$, $1$, and $N = 22$ lattice sites. Obviously, finite $M$
are visible as quasi-periodic oscillations. But the overall structure of
$C(\omega$) and the dc limit $C(\omega \to 0)$ are already well converged
for $M=200$.

(b) Similar to the typicality approach discussed before, the
statistical error of the Lanczos procedure is
\begin{equation}
\sigma(\epsilon)\leq {\cal O}\left(\frac{1}{\sqrt{RZ^*}}\right)\,,
\end{equation}
where $R$ is the number of random pure states used for sampling and
$Z^*$ is the thermodynamic sum
\begin{equation}
Z^*= \sum_n e^{-\beta (E_n-E_0)} \, ,
\end{equation}
where $E_0$ is the ground-state energy. Note that, for $\beta = 0$ and
$\beta = \infty$, the thermodynamic sum $Z^*$ is equivalent to the
effective dimension $d_\text{eff}$ in Eq.\ (\ref{deff}). For any finite
$\beta$, however, $Z^* \approx \text{d}_\text{eff}$ since energies $E_n$
located in the middle of the spectrum are approximately correct within
the Lanczos technique. We find $Z^* \approx 450$ for $J_\perp / J_\parallel
= 1$, $N =22$, and $\beta \, J_\parallel = 1$ and sample over $R = 10$
random pure states for the $\beta \neq 0$ cases in Fig.\ 5 our
Letter. For all $\beta = 0$ cases, $R = 1$.

In Fig.\ \ref{Fig10} we depict FTLM results for two different random
states $| \Phi \rangle_1 \neq | \Phi \rangle_2$ ($R = 1$) and an average
over several  $| \Phi \rangle_i$ ($R = 10$) for strong coupling $J_\perp
/ J_\parallel = 1$, two temperatures $\beta \, J_\parallel = 0$, $1$,
and $N = 22$ lattice sites. It is evident that the dependence on $R$
is negligibly small.

For a more detailed description of the implementation of FTLM and MCLM,
we refer the interested reader to Refs. \onlinecite{Sprelovsek2013,
SJaklic1994, SLong2004}.

\section{Equivalence of Ensembles}

\begin{figure}[t]
\includegraphics[width=0.9\columnwidth]{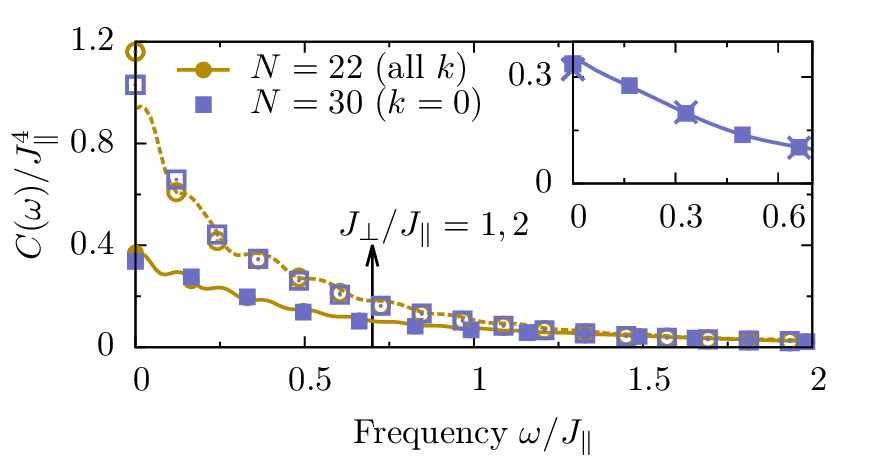}
\caption{(Color online) The same as Fig.\ 2 (b) of our Letter but
for the grand-canonical ensemble $\langle S^z \rangle = 0$.}
\label{Fig11}
\end{figure}

\begin{figure}[b]
\includegraphics[width=0.9\columnwidth]{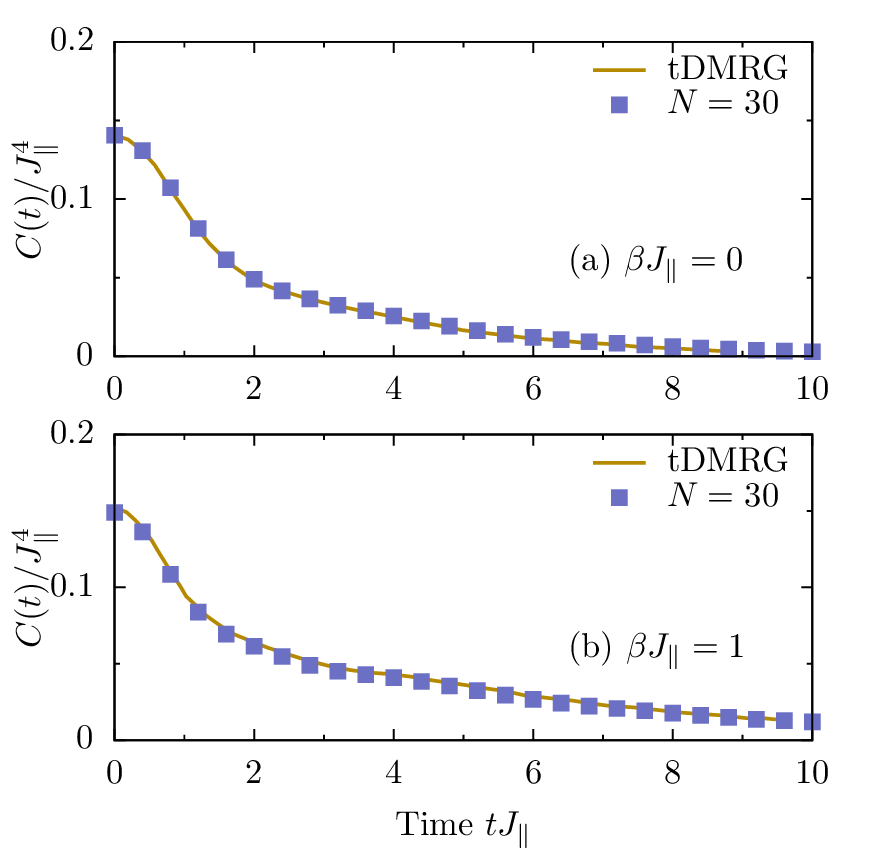}
\caption{(Color online) (a) Real-time data underlying the $J_\perp/
J_\parallel = 1$ and $N=30$ DQT spectrum in Fig.\ \ref{Fig11}. (b)
The same as (a) but for $\beta J_\parallel = 1 \neq 0$. For comparison,
in (a) and (b) the tDMRG data of Ref.\ \onlinecite{Snote} is depicted.}
\label{Fig12}
\end{figure}

Finally, we also demonstrate that all results presented in our Letter do
not depend on the restriction to the magnetization sector $S^z = 0$
(canonical ensemble). To this end, we repeat the calculation in Fig.\
2 (b) of our Letter for $\langle S^z  \rangle = 0$ (grand-canonical
ensemble), taking into account all magnetization sectors. The result
of this calculation is depicted in Fig.\ \ref{Fig11} and proves that
$S^z = 0$ and $\langle S^z \rangle = 0$ yield the same physics.

For the $J_\parallel/J_\perp = 1$ and $N=30$ DQT spectrum shown in Fig.\
\ref{Fig11}, we also depict in Fig.\ \ref{Fig12} (a) the underlying
real-time data. Furthermore, we compare this real-time data to the tDMRG
data of Ref.\ \onlinecite{Snote} and find excellent agreement at $\beta
J_\parallel = 0$. As illustrated in Fig.\ \ref{Fig12} (b), the agreement
between DQT and tDMRG data is also very good for $\beta J_\parallel
= 1$. Recall that the Fourier transforms in the inset of Fig.\ \ref{Fig11}
rely on much longer times than those times depicted in Fig.\ \ref{Fig12}.

\begin{figure}[t]
\includegraphics[width=0.9\columnwidth]{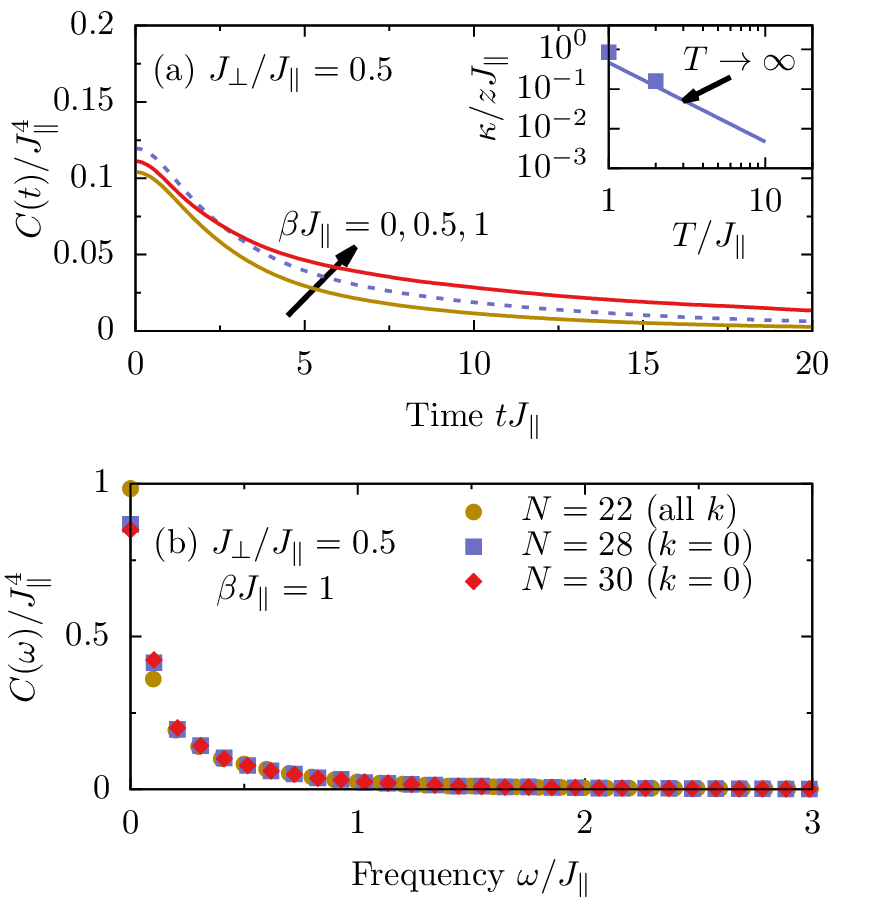}
\caption{(Color online) Real-time decay of the energy-current
autocorrelation function $C(t)$ for $\beta J_\parallel = 0, 0.5, 1$,
obtained from DQT for $J_\perp/J_\parallel = 0.5$ and $N = 30$. (b)
Spectrum for $\beta J_\parallel = 1$, obtained by Fourier transforming
data for finite times $t \leq 5 \tau \sim 30/J_\parallel$. Inset:
Temperature dependence of the conductivity $\kappa$, calculated by
$N=30$ DQT. The overall figure is similar to Fig.\ 5 of our Letter,
where $J_\perp/J_\parallel = 1$.}
\label{Fig13}
\end{figure}

\section{Temperature Dependence}

To illustrate that the temperature dependence of the heat conductivity
does not depend on the specific point $J_\perp / J_\parallel = 1$
considered in our Letter, we also repeat the calculation in Fig.\ 5
for $J_\perp / J_\parallel = 0.5$. The results of this calculation
are shown in Fig.\ \ref{Fig13}. Most importantly, we again find the
scaling $\kappa \propto T^{-2}$ down to $T$ on the order of the
exchange coupling, as shown in the inset of Fig.\ \ref{Fig13}
(a).

\newpage


\begin{thebibliography}{99}

\bibitem{johnston2000}
D. C. Johnston {\it et al.},
Phys. Rev. B \textbf{61}, 9558 (2000).

\bibitem{deutsch1991} J. M. Deutsch,
Phys. Rev. A {\bf 43}, 2046 (1991).

\bibitem{srednicki1994} M. Srednicki,
Phys. Rev. E {\bf 50}, 888 (1994).

\bibitem{rigol2008} M. Rigol, V. Dunjko, and M. Olshanii,
Nature {\bf 452}, 854 (2008).

\bibitem{trotzky2008}
S. Trotzky {\it et al.},
Science \textbf{319}, 295 (2008).

\bibitem{gambardella2006}
P. Gambardella,
Nature Mat. {\bf 5}, 431 (2006).

\bibitem{kruczenski2004}
M. Kruczenski,
Phys. Rev. Lett. {\bf 93}, 161602 (2004).

\bibitem{kim1996} Y. B. Kim,
Phys. Rev. B {\bf 53}, 16420 (1996).

\bibitem{zotos1997} X. Zotos, F. Naef, and P. Prelov\v{s}ek,
Phys. Rev. B {\bf 55}, 11029 (1997).

\bibitem{kluemper2002} A. Kl\"umper and K. Sakai,
J. Phys. A: Math. Gen. {\bf 35}, 2173 (2002).

\bibitem{sologubenko2000}
A. V. Sologubenko {\it et al.},
Phys. Rev. Lett. {\bf 84}, 2714 (2000).

\bibitem{hess2001}
C. Hess {\it et al.},
Phys. Rev. B {\bf 64}, 184305 (2001).

\bibitem{hess2007} C. Hess {\it et al.}
Phys. Rev. Lett. {\bf 98}, 027201 (2007).

\bibitem{hlubek2010} N. Hlubek {\it et al.},
Phys. Rev. B {\bf 81}, 20405R (2010).

\bibitem{shastry1990} B. S. Shastry and B. Sutherland,
Phys. Rev. Lett. {\bf 65}, 243 (1990).

\bibitem{narozhny1998} B. N. Narozhny, A. J. Millis, and N. Andrei,
Phys. Rev. B {\bf 58}, 2921R (1998).

\bibitem{zotos1999} X. Zotos,
Phys. Rev. Lett. {\bf 82}, 1764 (1999).

\bibitem{benz2005} J. Benz {\it et al.},
J. Phys. Soc. Jpn. {\bf 74}, 181 (2005).

\bibitem{fujimoto2003} S. Fujimoto and N. Kawakami,
Phys. Rev. Lett. {\bf 90}, 197202 (2003).

\bibitem{prosen2011} T. Prosen,
Phys. Rev. Lett. {\bf 106}, 217206 (2011).

\bibitem{prosen2013} T. Prosen and E. Ilievski,
Phys. Rev. Lett. {\bf 111}, 057203 (2013).

\bibitem{herbrych2011} J. Herbrych, P. Prelov\v{s}ek, and X. Zotos,
Phys. Rev. B {\bf 84}, 155125 (2011).

\bibitem{karrasch2012} C. Karrasch, J. H. Bardarson, and J. E. Moore,
Phys. Rev. Lett. {\bf 108}, 227206 (2012).

\bibitem{karrasch2013-1} C. Karrasch {\it et al.},
Phys. Rev. B {\bf 87}, 245128 (2013).

\bibitem{steinigeweg2014-1} R. Steinigeweg, J. Gemmer, and W. Brenig,
Phys. Rev. Lett. {\bf 112}, 120601 (2014).

\bibitem{steinigeweg2014-2} R. Steinigeweg, J. Gemmer, and W. Brenig,
Phys. Rev. B {\bf 91}, 104404 (2015).

\bibitem{carmelo2014} J. M. P. Carmelo, T. Prosen, and D. K. Campbell,
preprint, arXiv:1407.0732 (2014).

\bibitem{sirker2009} J. Sirker, R. G. Pereira, and I. Affleck,
Phys. Rev. Lett. {\bf 103}, 216602 (2009).

\bibitem{sirker2011} J. Sirker, R. G. Pereira, and I. Affleck,
Phys. Rev. B {\bf 83}, 035115 (2011).

\bibitem{grossjohann2010} S. Grossjohann and W. Brenig,
Phys. Rev. B {\bf 81}, 012404 (2010).

\bibitem{znidaric2011} M. \v{Z}nidari\v{c},
Phys. Rev. Lett. {\bf 106}, 220601 (2011).

\bibitem{steinigeweg2011} R. Steinigeweg and W. Brenig,
Phys. Rev. Lett. {\bf 107}, 250602 (2011).

\bibitem{karrasch2014} C. Karrasch, J. E. Moore, and F. Heidrich-Meisner,
Phys. Rev. B {\bf 89}, 075139 (2014).

\bibitem{thurber2001} K. R. Thurber {\it et al.},
Phys. Rev. Lett {\bf 87}, 247202 (2001).

\bibitem{maeter2012} H. Maeter {\it et al.},
J. Phys.: Condens. Matter {\bf 25}, 365601 (2013).

\bibitem{ronzheimer2013} J. P. Ronzheimer {\it et al.},
Phys. Rev. Lett. {\bf 110}, 205301 (2013).

\bibitem{hild2014} S. Hild  {\it et al.},
Phys. Rev. Lett. {\bf 113}, 147205 (2014).

\bibitem{xiao2014} F. Xiao {\it et al.},
preprint, arXiv:1406.3202 (2014).

\bibitem{shimshoni2003} E. Shimshoni, N. Andrei, and A. Rosch,
Phys. Rev. B {\bf 72}, 059903 (2005).

\bibitem{rozhkov2005} A. V. Rozhkov and A. L. Chernyshev,
Phys. Rev. Lett. {\bf 94}, 087201 (2005).

\bibitem{hlubek2012} N. Hlubek {\it et al.},
J. Stat. Mech. {\bf 12}, P03006 (2012).

\bibitem{huang2013} Y. Huang, C. Karrasch, J. E. Moore,
Phys. Rev. B {\bf 88}, 115126 (2013).

\bibitem{karrasch2013-2} C. Karrasch, R. Ilan, and J. E. Moore,
Phys. Rev. B {\bf 88}, 195129 (2013).

\bibitem{karahalios2009} A. Karahalios {\it et al.},
Phys. Rev. B {\bf 79}, 024425 (2009).

\bibitem{heidrichmeisner2003} F. Heidrich-Meisner {\it et al.},
Phys. Rev. B {\bf 68}, 134436 (2003).

\bibitem{steinigeweg2013} R. Steinigeweg, J. Herbrych, P. Prelov\v{s}ek,
Phys. Rev. E {\bf 87}, 012118 (2013).

\bibitem{Elbio1996} E. Dagotto and T. M. Rice,
Science {\bf 271}, 618 (1996).

\bibitem{Bella2010a} B. Lake {\it et al.},
Nature Phys. {\bf 6}, 50 (2010).

\bibitem{Notbohm2007} S. Notbohm {\it et al.},
Phys. Rev. Lett. {\bf 98}, 027403 (2007). 

\bibitem{Schmidiger2013} D. Schmidiger {\it et al.},
Phys. Rev. B {\bf 88}, 094411 (2013).

\bibitem{Thielemann2009a} B. Thielemann {\it et al.},
Phys. Rev. Lett. {\bf 102}, 107204 (2009).

\bibitem{Thielemann2009b} B. Thielemann {\it et al.},
Phys. Rev. B {\bf 79}, 020408 (2009).

\bibitem{George1997a} G. B. Martins, M. Laukamp, J. Riera, and E. Dagotto,
Phys. Rev. Lett. {\bf 78}, 3563 (1997).

\bibitem{Tranquada2004} J. M. Tranquad {\it et al.},
Nature {\bf 429}, 534 (2004).

\bibitem{Elbio1992} E. Dagotto, J. Riera, and D. Scalapino,
Phys. Rev. B {\bf 45}, 5744(R) (1992).

\bibitem{Garcia2004} J. J. Garc\'ia-Ripoll, M. A. Martin-Delgado, and J. I. Cirac,
Phys. Rev. Lett. {\bf 93}, 250405 (2004).

\bibitem{Ying2005} Y. Li {\it et al.},
Phys. Rev. A {\bf 71}, 022301 (2005).

\bibitem{Deshpande2009} V. V. Deshpande {\it et al.},
Science {\bf 323}, 106 (2009).

\bibitem{jung2006} P. Jung, R. W. Helmes, and A. Rosch,
Phys. Rev. Lett. {\bf 96}, 067202 (2006).

\bibitem{jung2007} P. Jung and A. Rosch,
Phys. Rev. B {\bf 76}, 245108 (2007).

\bibitem{zotos2004} X. Zotos,
Phys. Rev. Lett. {\bf 92}, 067202 (2004).

\bibitem{note} C. Karrasch, D. M. Kennes, and F. Heidrich-Meisner,
Phys. Rev. B {\bf 91}, 115130 (2015).

\bibitem{steinigeweg2014-3} R. Steinigeweg {\it et al.},
Phys. Rev. B {\bf 90}, 094417 (2014).

\bibitem{elsayed2013} T. A. Elsayed and B. V. Fine,
Phys. Rev. Lett. {\bf 110}, 070404 (2013).

\bibitem{gemmer2003} J. Gemmer and G. Mahler,
Eur. Phys. J. B {\bf 31}, 249 (2003).

\bibitem{goldstein2006} S. Goldstein {\it et al.},
Phys. Rev. Lett. {\bf 96}, 050403 (2006).

\bibitem{popescu2006} S. Popescu, A. J. Short, and A. Winter,
Nature Phys. {\bf 2}, 754 (2006).

\bibitem{reimann2007} P. Reimann,
Phys. Rev. Lett. {\bf 99}, 160404 (2007).

\bibitem{white2009} S. R. White,
Phys. Rev. Lett. {\bf 102}, 190601 (2009).

\bibitem{bartsch2009} C. Bartsch and J. Gemmer,
Phys. Rev. Lett. {\bf 102}, 110403 (2009).

\bibitem{bartsch2011} C. Bartsch and J. Gemmer,
EPL {\bf 96}, 60008 (2011).

\bibitem{sugiura2012} S. Sugiura and A. Shimizu,
Phys. Rev. Lett. {\bf 108}, 240401 (2012).

\bibitem{hams2000} A. Hams and H. De Raedt,
Phys. Rev. E {\bf 62}, 4365 (2000).

\bibitem{steinigeweg2010}  R. Steinigeweg and R. Schnalle,
Phys. Rev. E {\bf 82}, 040103R (2010).

\bibitem{steinigeweg2011-1} R. Steinigeweg,
Phys. Rev. E {\bf 84}, 011136 (2011).

\bibitem{prelovsek2013} A recent review is given in: P. Prelov\v{s}ek and J. Bon\v{c}a,
{\it Ground State and Finite Temperature Lanczos Methods} in {\it Strongly Correlated Systems}, Solid-State Sciences 176 (Springer, Berlin, 2013).

\bibitem{SM} See Supplemental Material.

\end{thebibliography}

\begin{thebibliography}{99}

\bibitem{Sjung2006} P. Jung, R. W. Helmes, and A. Rosch,
Phys. Rev. Lett. {\bf 96}, 067202 (2006).

\bibitem{Sjung2007} P. Jung and A. Rosch,
Phys. Rev. B {\bf 76}, 245108 (2007).

\bibitem{Ssteinigeweg2010}  R. Steinigeweg and R. Schnalle,
Phys. Rev. E {\bf 82}, 040103R (2010).

\bibitem{Ssteinigeweg2011-1} R. Steinigeweg,
Phys. Rev. E {\bf 84}, 011136 (2011).

\bibitem{Selsayed2013} T. A. Elsayed and B. V. Fine,
Phys. Rev. Lett. {\bf 110}, 070404 (2013).

\bibitem{Ssteinigeweg2014-1} R. Steinigeweg, J. Gemmer, and W. Brenig,
Phys. Rev. Lett. {\bf 112}, 120601 (2014).

\bibitem{Ssteinigeweg2014-2} R. Steinigeweg, J. Gemmer, and W. Brenig,
Phys. Rev. B {\bf 91}, 104404 (2015).

\bibitem{Sprelovsek2013} P. Prelov\v{s}ek and J. Bon\v{c}a,
{\it Ground State and Finite Temperature Lanczos Methods} in {\it Strongly Correlated Systems}, Solid-State Sciences 176 (Springer, Berlin, 2013).

\bibitem{SJaklic1994} J. Jakli\v{c} and P. Prelov\v{s}ek,
Phys. Rev. B {\bf 49}, 5065(R) (1994).

\bibitem{SLong2004} M. W. Long, P. Prelov\v{s}ek, S. El Shawish, J. Karadamoglou, and X. Zotos,
Phys. Rev. B {\bf 70}, 205129 (2004).

\bibitem{Snote} C. Karrasch, D. M. Kennes, and F. Heidrich-Meisner,
Phys. Rev. B {\bf 91}, 115130 (2015).

\end{thebibliography}
\end{document}